\documentclass[12pt]{article}
\usepackage{makeidx}  % allows for indexgeneration
\usepackage{graphicx} 

\usepackage{url}

\begin{document}

\title{Doubleclick Ad Exchange Auction\footnote{Work done at Google, Inc.}}
\author{Yishay Mansour~\footnote{Tel Aviv Univ., \em{mansour@tau.ac.il}}, S. Muthukrishnan~\footnote{Rutgers Univ., {\em muthu@cs.rutgers.edu}}, and Noam Nisan~\footnote{Hebrew Univ., {\em noam@cs.huji.ac.il}}}
\date{}
\maketitle              % typeset the title of the contribution

\begin{abstract}
Display advertisements on the web are sold via ad exchanges that use real time auction. 
We describe the challenges of designing a suitable auction, and present a simple auction called
the {\em Optional Second Price} (OSP) auction that is currently used in Doubleclick Ad Exchange. 
\end{abstract}

\section{Introduction to Web Display Ads}
Web is a pervasive entity: millions of publishers produce content and configure webpages; 
hundreds of millions of users worldwide browse the web every day and access these pages; as a result, web has become a powerful communication medium. Its economics is predominantly driven by advertisers who wish to get the attention of the users using the publishers as the channel for placing advertisements (ads) on the pages.
Where in the webpage and how many ads are shown are determined by the publisher of the page. Which ads are shown is
determined by offline negotiation between sales teams of publishers and advertisers, perhaps via intermediaries like
ad agencies and networks. How these ads are actually delivered from advertiser to the web user is achieved by 
ad delivery systems that are sophisticated computer and overlay communication systems on the Internet. Payment depends on
whatever sales teams negotiate, and may vary from advertiser to advertiser to reach the same user at the same page. 
There is little transparency.

An emerging way of selling and buying ads on the Internet is via an
{\em exchange} that brings sellers (publishers) and buyers (advertisers) together to
a common, automatic marketplace.
There are exchanges in the world for trading financial securities to currency, physical goods, virtual credits, and much more. 
Exchanges serve many purposes from
bringing efficiency, to eliciting prices, generating capital, aggregating information etc. 
%Market Microstructure is the area that studies all aspects of such exchanges. 
Ad exchanges are recent.
%Ad exchanges offer ad networks and publishers to transact centrally for ads. 
RightMedia~\cite{RM}, adBrite~\cite{adbrite}, OpenX~\cite{openx}, and DoubleClick~\cite{AdX} 
are examples.  
\begin{itemize}
\item
Publishers expect to get the best price from the exchange, 
better than from any specific ad network; in addition, publishers get liquidity. 
\item
Advertisers
get access to a large inventory at the exchange, and in addition, the ability to 
target more precisely across web pages. 
\item
Finally, the exchange is a clearing house ensuring the flow of money. 
\end{itemize}
In many ways, these ad exchanges are modeled after financial
stock exchanges. Since 2005 when RightMedia appeared, ad exchanges have become popular. In Sept 2009,
RightMedia averaged 9 billion transactions a day with 100's of thousands of buyers and sellers. DoubleClick's  ad exchange has been active since 2009. It seems ad 
exchanges are likely to become a major platform for trading ads. 
%They represent an elaborate ``ecosystem'' with many parties, each optimizing and 
% strategizing on their own behalf and on behalf of their customers. 

\section{How Ad Exchanges Work} 
There are several key decisions that ultimately determine the architecture of ad exchanges: these decisions include, for example, what goods should be traded, what should be the relationship among ad networks that exchanges endogenize, what are informational interfaces including bids, minimum prices, and winner/loser notification, and what are the contracts between the exchange and publishers or ad networks etc.
%As an example, 
%There are a number of decisions involved in designing ad marketplaces. 
For example, what commodities
should be traded? One can imagine trading contracts for bulk impressions (eg., 1M
impressions per day in YouTube homepage for a movie trailer). Instead, 
ad exchanges trade individual impressions. More sophisticated
contracts can be crafted on top of this spot market. Another issue is who will 
be the participants in the market? Much like financial exchanges that only let licensed
brokers trade, ad exchanges let ad networks trade on the exchange on behalf on individual
advertisers. 

The {\em AdX} model in~\cite{Muthu} is an abstraction of ad exchanges. 
% for ad marketplaces. 
We describe it here in a suitable form as the basis for our discussions (more details can be found 
in~\cite{Muthu}). It is defined as a sequence of events. 
\begin{enumerate}
\item
User $u$ visits the webpage $w$ of publisher $p(w)$ that has, say, a single slot for ads. 

\item
Publisher $p(w)$ contacts the exchange $E$ with $(w, P(u),\rho)$ where $\rho$ is 
the {\em minimum price} $p(w)$ is willing to take for the slot in $w$, and $P(u)$ is the information about user $u$ that $P(w)$ shares with $E$. 

\item
The exchange $E$ contacts ad networks $a_1,\ldots,a_m$ with 
$(E(w), E(u))$, where $E(w)$ is information about $w$ provided by
$E$, and $E(u)$ is the information about $u$ provided by $E$. $E()$ may be 
potentially  different
from $P(u)$. 
%, and likewise $E(\rho)$,
%information about $\rho$ provided by $E$. 

\item
Each ad network $a_i$ returns $(b_i, d_i)$ on behalf of its customers which are the
advertisers; $b_i$ is its {\em 
bid}, that is, the maximum it is willing to pay for the slot in page $w$ and 
$d_i$ is the ad it wishes to be shown. The ad networks may also choose not to
return a bid. 

\item
Exchange $E$ determines a winner $i^*$ for the ad slot among all $(b_i, d_i)$'s
and its price  $c_{i^*} $,    $\rho\leq c_{i^*} \leq b_{i^*}$ via an {\em auction}.  This auction will 
discussed in detail later. 

\item
Exchange $E$ returns winning ad 
$d_{i^*}$ to $p(w)$ and price $c_{i^*}$ to $i^*$. 

\item
The publisher $p(w)$ serves webpage $w$ with ad $d_{i^*}$ to user $u$. 
This is known as an {\em impression} of ad $d_{i^*}$.
\end{enumerate}

\begin{center}
\begin{figure}
\includegraphics[height=60mm,width=150mm]{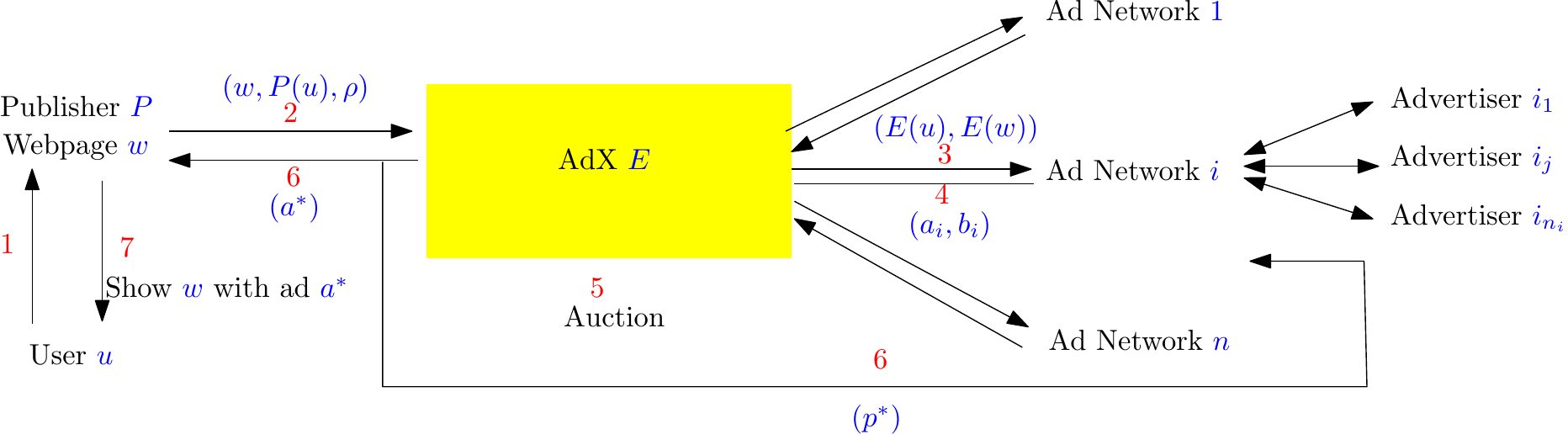}
\caption{AdX}
\label{fig:adx}
\end{figure}
\end{center}

The flow is shown in Figure~\ref{fig:adx}. It is important for our discussion to realize that the entire flow occurs between the time $u$ requests page $w$, and $w$ is shown to the user. Hence, the process needs to take a few 100 msec and no longer. Second, potentially, every view on the Internet may generate a call to the Exchange, so potentially 10's of billions of transactions may result in the Exchange in a day. Finally, Exchanges are potentially the gateway for the Billions of dollars of the display ads Economy that underpins the Web, and they modulate the flow from millions of advertisers to millions of content generators. As a result, there are many high performance engineering requirements of the Exchange and optimizations, which we do not address here. Also, there are significant reasons for the various informational interfaces provided in the Exchange which we do not discuss 
here~\cite{Muthu}. 

\section{The Problem of Auctions at AdX}
At the core of AdX is an auction to determine the winner for the impression of an ad.  There are 
potentially other auctions, in some cases, at the 
publisher or at the ad networks to determine the winner and bidders resp. which we do not directly emphasize here. The AdX auction governs the short term revenues of  three parties directly:
\begin{itemize}
\item
the publisher: higher price translates directly to higher revenue, 
\item
the ad exchange: higher price potentially leads to higher commissions for transacting, and 
\item
perhaps surprisingly, even the 
ad networks:  higher price leads to potentially higher commissions for handling ad campaigns on behalf of advertisers.
\end{itemize} 
There are also long term revenue concerns. Making the auction more efficient, incentive-compatible, transparent, stable and fair will ultimately increase the adoption of this ad market by buyers and sellers and increase revenues long term. 

At a high level, the auction problem at the exchange seems the simple case of single item auction. Decades of research in Auction Theory
~\cite{Krishna} have identified efficient (VCG) auctions, Myerson's revenue-optimal auctions~\cite{Myerson}, \footnote{Both of these auctions have been recognized with Nobel prizes.}
and developed a structural theory of equivalence between certain auction types for single item auctions. On closer inspection, the auction problem at AdX is not simple. Here are two example issues:
\begin{itemize}
\item
{\em Intermediaries.} Intermediaries on buy (and also sell) side influence the information available to the auctioneer and distort the market. As an example, say Network 1 has two advertisers who bid 10 and 8; Network 2 has one advertiser with bid 5. From the description above, it will 
follow that AdX only sees bids 10 and 5 and hence, Network 1 wins the good at price 5, while it can charge 8 to 
its top bidder using second price! The observation is that due to the presence of networks, 
AdX does not have  the full information and can not implement {\em true} second price on the entire book of bids. Further,  the exchange
loses information about all the bids, and hence, loses the ability to shift the potential revenue away from the networks if so desired. 
Readers can imagine other auctions at the exchange and realize that this problem persists when networks strategize. 

\item
{\em Alternative Channels.} Publishers set min price $\rho$ to optimize their revenue and other considerations. On top of that,  it is expected that AdX should impose additional reserve price as Myerson's theory~\cite{Myerson} will indicate in order to maximize its revenue.  In a simple formulation of this problem~\cite{Pai}, it is shown that this reserve will tend to $0$ as number of intermediaries 
increases, but the theory is unclear about the impact of a reserve price at AdX. If AdX imposes its additional reserve price,  the goods will remain unsold by AdX for a fraction of the times it would have sold otherwise. Hence, its {\em fill rate}, that is, the  percentage of times it returns an ad to the publisher will go down. This is an important metric of the health of AdX and if fill rate declines, 
the publisher may choose not to come to AdX or fill using alternative sales channels. Thus, AdX may not realistically be able to strategize and compete against alternative channels if they wanted to ensue a vibrant market.
\end{itemize}
Therefore, auction design at AdX is a challenge. 

\section{Our Auction Solution} 
One option is for the AdX to adopt first price aucton. Observe that this is implementable 
{\em truly}  in AdX in presence of networks because no information about any bid other than the 
highest bid at any network is needed to implement it. However, first price auction, even without
intermediaries, has well known strategic problems of instability. Thus, it is desirable to keep 
the second price logic. 
How to enforce it is a difficult business decision. One approach  is to change the bidding language 
and ask each network to 
reveal the entire book of bids, but networks can strategize and invalidate this approach. One might even consider using legal contracts to make networks comply. This will be a big business hurdle, and jeopardize the adaptation of AdX. AdX might reveal the winning price to end advertisers for each 
impression (bypassing the networks) and thereby shift the savings to them. This does not satisfy the publishers by getting the 
second price revenue from the entire book of bids and additionally this 
alienates the networks which are formidable players in this ecosystem.

Our final solution is exceedingly simple, as indeed it needs to be given  performance constraints. 
\begin{itemize}
\item
Each network $i$  that participates submits {\em mandatory} bid $b_i$ and 
an {\em optional} bid $o_i\leq b_i$ (which could be $0$). 
\item
AdX runs a second price auction, using
publisher-specified $\rho$ as the reserve price, and charges winner $i$, 
$\max\{\max_{j \not= i} b_j, o_i,\rho\}$. 
\end{itemize}

We call this the {\em Optional Second Price} (OSP) Auction. 

\section{Properties of the OSP Auction}

The OSP auction represents  many business decisions. 
First, we ignore short term revenue of AdX and do not introduce any reserve price on top of publisher specified $\rho$. We let publishers control 
their revenue with min price $\rho$ that they strategize, exogenous to AdX. 
We rejected disruptive solutions like making prices public after each auction to advertisers, and let networks handle relationships with advertisers. This in particular would mean, scenario above of a network winning a good at 5 for 10 bid is feasible. Networks are an important part of the display
ads business: in some cases, they produce content and plan campaign smartly, and are already entrenched in the ecosystems. So, AdX auction lets them implement their strategic goals, no matter if it is not a true second price auction at the AdX. 

On the other hand,  the OSP auction  has certain desirable properties:

\medskip
\noindent
{\bf Property P1.} 
If a network $i$, wishes to simulate second price auction among {\em its} bidders, it can do so using the optional bid $o_i$ by correctly specifying 
the second highest bid among its bidders as $o_i$  ($=8$ in the example).  In this case, from the point of view of all its bidders,
the OSP auction is completely indistinguishable 
from a true second price auction.  This is true whether or not any of the
other networks send the correct second bid.
Thus, honest second price networks can simulate their business with little overhead. 
In the absence of this feature, networks that have contractual obligation with their bidders to run a second price auction 
(such as Google's Adwords network) will have to significantly modify their logging and billing system to correctly track the spend. 
Further, they will have to explain to their customer with the second highest bid why they did not win the impression that was sold at  
a price lower than their bid. Thus their business as well as technology problems are immediately solved by using the optional bid feature honestly. 

\medskip
\noindent
{\bf Property P2.}
There are other possible options for a networks $i$ regarding the auction it implements among its bidders.  Here are some natural ones:
(1 ) It can declare and simulate a second price auction to its bidders, pocketing the difference 
$(\max_{k} b_{i_k}) - (\max_{j \not= i} b_j)$
whenever it controls both the highest and second highest bids, where $b_{i_k}$'s denote the bids of its bidders.  In this case, from the point of view of all its bidders, the OSP auction is still indistinguishbable 
from a true second price auction, and strategic bidders will be truthful.  (2) It may choose to give back all the savings to the bidders, i.e. charge its winning
bidders the price of $\max_{j \not= i} b_j$.  This is similar to a bidding club, and in this case strategic 
bidders should not be truthful but rather will tend to {\em over-bid}.  Underbidding is weakly dominated by truth. (3) It may charge a fixed price of its choice 
from all its bidders, rather than conducting an auction (this is a common strategy for ad networks).  In this case only 
impressions where this fixed price is higher than $\max_{j \not= i} b_j$
will be sold to the nework's bidders and some network policy must dictate who wins in case of multiple advertisers that are willing to pay the fixed price.  (4)
A first price auction among it's bidders is a possibility, in which case $\max_{j \not= i} b_j$ serves as a reserve price, and we expect strategic bidders to 
slightly underbid.

\medskip
\noindent
{\bf Property P3.} The expected gain of a network from not reporting the correct second price which is equal to the expected loss of the publisher is small, as
long as the network is not too large a fraction of the market. 
We will present a simple analysis to quantify how much lying about the optional bid profits
a network that pretends to be a second price auction and pockets the difference -- the first option above\footnote{This will also be an upper bound
on the gains of the bidders of a network that passes on the gains to them -- the second option -- as in this case strategic bidders never underbid.}. 
Say there are $k$ networks,  $k\geq 2$, and  bidders with arbitrary bids 
$d_1 \geq \ldots  \geq d_n$ are independently, uniformly, randomly assigned to the networks.  Say network $1$ 
has the highest bid  $d_1$. 
There are two cases. 
\begin{itemize}
\item
Network $1$ is truthful about $o_1$ where $o_1$ is the second highest bid among bidders assigned to network $1$, that is,
$o_1=\max_{k} b_{1_k}$. 
All networks are truthful about mandatory bid $b_i$'s.  It is immediate that price of $1$ is   
$d_2$, no matter the assignment of bidders to networks.

\item
Network $1$ is not truthful about $o_1$. Then, its price is $d_2$ if the bidder with bid $d_2$ was represented by one of the networks other than $1$ 
which happens with probability $1-1/k$; its price is $d_3$ if $d_2$ is assigned to network $1$ and $d_3$ is assigned to one of other networks, which 
happens with probability $(1/k)(1-1/k)$; continuing like this, network $1$'s price is $d_j$ with probability $ (1/k^{j-2})(1-1/k)$.  
In other words the gain of network $1$ is 
$$\sum_{j=2}^n \frac{d_j}{k^{j-2}} \left(1-\frac{1}{k}\right) - d_2 
=
\sum_{i=1}^n \frac{d_{i+2}-d_{i+1}}{k^i},$$ 
and this is the publisher's loss.

We can first observe that for any values of the $d_i$'s the loss of the publisher 
is bounded from above by $d_2/k$, which is a loss of at most $1/k$ fraction of the revenue that would have been obtained with the correct reporting
of the second price.  This is the loss conditioned on winning network not reporting the second price correctly.  
When $t$ out of the $k$ networks do not report their second price correctly, and the bidders choosing between them uniformly, then 
the expected loss, is at most $t/k^2$ fraction of revenue.

The analysis above holds for any values of the $d_i$'s, and is tight only when $d_3=d_4=...=d_n=0$.  Let us consider a more typical scenario, where the $d_i$'s
are, e.g., chosen uniformly at random in $[0,1]$.  In this case, the expected value of $d_2$ is $1-2/(n+1)$ and the expected value of $d_{i+2}-d_{i+1}$ is $1/(n+1)$.
It follows that the expected loss if the wining network lies is 
$$\sum_{i=1}^n \frac{1}{(n+1)k^i},$$ 
which is $O(1/(kn))$ fraction of the revenue.
\end{itemize}

\noindent
The analyses above assume the bidders do not differentiate among networks. We leave it open to analyze the OSP auction under more 
sophisticated models of how bidders gravitate towards different networks 
(eg., a good model might be {\em proportional} where bidders join networks with probability proportional to their current size).

\section{Use of the OSP Auction}

Google's DoubleClick Ad Exchange uses our OSP auction~\cite{faq}. It has been operational worldwide since Sept 2009. The number of transactions at this exchange (ie, transactions that use our auction) exceeds the number of transactions at financial exchanges worldwide and this is a conservative benchmark. Discussion of a year of experience with this exchange is in Google blog~\cite{blog}. Our optional bid feature is being used (among other things) to let Google's Adwords network bid via Ad Exchange~\cite{faq}, where this feature is used to simulate AdWord's native second price auction seamlessly, without additional effort. This is crucially due to property P1 above. 
It is rare that three of us researchers were involved at the beginning when DoubleClick Ad Exchange was  built, and had the opportunity to impact the auction. 

\section{Concluding Remarks}
AdX makes the well-trenched display ads business based on offline negotiations, to be more automatic, market-driven and sustains a very large ecosystem on buy and sell sides. We have demonstrated that 
even in single item auction as it arises in AdX, there are nontrivial design issues due to the presence of intermediary networks. We have proposed the Optional Second Price (OSP) auction which has desirable nuances and is currently used in AdX. 

\bibliographystyle{plain}

\end{document}